\documentclass[12pt,preprint]{emulateapj}

\def\ltsima{$\; \buildrel < \over \sim\;$}
\def\ltsim{\lower.5ex\hbox{\ltsima}}
\def\gtsima{$\; \buildrel > \over\sim \;$}
\def\gtsim{\lower.5ex\hbox{\gtsima}}
\def\ms{$M_{\odot}$ }
\def\msp{$M_{\odot}$}

\begin{document}

\title{The \lowercase{r}-process in magnetorotational supernovae}

\author{Takuji Tsujimoto\altaffilmark{1} and Nobuya Nishimura\altaffilmark{2}}

\affil{
\altaffilmark{1}{National Astronomical Observatory of Japan, Mitaka-shi, Tokyo 181-8588, Japan; taku.tsujimoto@nao.ac.jp}\\
\altaffilmark{2}{Astrophysics Group, Keele University, ST5 5BG Keele, UK}
}

\begin{abstract}
One of the hottest open issues involving the evolution of r-process elements is fast enrichment in the early Universe. Clear evidence for the chemical enrichment of r-process elements is seen in the stellar abundances of extremely metal poor stars in the Galactic halo. However, small-mass galaxies are the ideal testbed for studying the evolutionary features of r-process enrichment given the potential rarity of production events yielding heavy r-process elements. Their occurrences become countable and thus an enrichment path due to each event can be found in the stellar abundances. We examine the chemical feature of Eu abundance at an early stage of ${\rm[Fe/H]} \lesssim -2$ in the Draco and Sculptor dwarf spheroidal (dSph) galaxies. Accordingly, we constrain the properties of the Eu production in the early dSphs. We find that the Draco dSph experienced a few Eu production events, whereas Eu enrichment took place more continuously in the Sculptor dSph due to its larger stellar mass. The event rate of Eu production is estimated to be about one per $100$--$200$ core-collapse supernovae, and a Eu mass of $\sim (1-2) \times 10^{-5}$\ms per single event is deduced by associating this frequency with the observed plateau value of ${\rm [Eu/H]} \sim -1.3$ for ${\rm [Fe/H]} \gtrsim -2$. The observed plateau implies that early Eu enrichment ceases at ${\rm [Fe/H]} \approx -2$. Such a selective operation only in low-metallicity stars supports magnetorotational supernovae, which require very fast rotation, as the site of early Eu production. We show that the Eu yields deduced from chemical evolution agree well with the nucleosynthesis results from corresponding supernovae models. 
\end{abstract}


\section{Introduction}

What is the astrophysical object producing r-process elements and how has r-process enrichment proceeded through cosmic time? Clear answers to these questions are still veiled. Recent updates from both observational and theoretical studies of the r-process site, in particular by the detection of a near-infrared light bump in the afterglow of a short-duration $\gamma$-ray burst (sGRB)\citep[e.g.,][]{Tanvir_13} and successful nucleosynthesis calculations \citep[e.g.,][]{Rosswog_14, Wanajo_14, Goriely_15}, have indicated neutron star (NS) mergers as the most promising astronomical site. Core-collapse supernovae (CC-SNe) driven by neutrino heating are another candidate for the r-process site, buy they are unable to synthesize heavy r-process elements such as Eu and Ba. The required conditions for r-process nucleosynthesis are not achieved in the proto-NS winds due to the strong effect of neutrino absorption \citep[e.g.,][]{Thompson_01, Fischer_10, Wanajo_13}. This fact has turned out to strengthen the NS merger origin scenario. 

New research on r-process synthesis has renewed interest in the Galactic chemical evolution (GCE) of r-process elements in terms of the enrichment taking place through NS mergers. Though previous researchers have pointed out the fatal problems with the notion of enrichment by NS mergers \citep{Mathews_90, Argast_04}, \citet{Tsujimoto_14} successfully reproduced an observed Eu feature in the Galaxy using a GCE model that considers a unique propagation of the ejecta of NS mergers in the scheme of hierarchical galaxy formation. They also found that an early release of Eu from (at least some fraction of) NS mergers with a short timescale of $\sim 10^7$ year is an inevitable condition for explaining the presence of very low-metallicity ([Fe/H]\ltsim $-3$) stars enriched by r-process elements. Such a prompt enrichment by NS mergers is also claimed by other GCE studies \citep{Matteucci_14, Ishimaru_15, Wehmeyer_15}, and can be a possible channel in population synthesis models \citep{Belczynski_06, Dominik_12}. However, the coalescence time for an NS merger depends critically on the binary evolution during a common envelope phase which is poorly understood. In addition, the above requirement is in contrast with a typical NS merger time of $3$--$4$ Gyr deduced from the analysis of {\it Swift} sGRB samples \citep{Wanderman_15}. 

\citet{Tsujimoto_14} have extended the discussion of the site of r-process elements to nearby dwarf spheroidal (dSph) galaxies. They have found compelling evidence for NS mergers as the origin of r-process elements, in the observed chemical feature of Eu in the Draco, Sculptor, and Carina dSphs; these dSphs exhibit a constant [Eu/H] of $\sim -1.3$ with no apparent increase in the Eu abundance over a  metallicity range of $-2$\ltsim [Fe/H]\ltsim$-1$. This implies  no Eu production events as more than $10^3$ CC-SNe increase the galactic Fe abundance. This level of rarity of Eu production is compatible with the frequency expected for NS mergers \citep{Tsujimoto_14}. The next question is how these dSphs which do not undergo any Eu production events for [Fe/H]\gtsim$-2$ are enriched up to  [Eu/H]$\sim -1.3$ until [Fe/H]$\sim -2$. \citet{Tsujimoto_15} examined Eu abundances for [Fe/H]$<-2$ in the Draco and Sculptor dSphs from a purely observational viewpoint and concluded that [Eu/H] remarkably increases from $<-4$ to $\sim -1.3$ inside these dSphs, including a conspicuous intermittent increase in the Draco dSph. These revealed Eu features suggest that the contributor of Eu in the early dSphs is active only in the low-metallicity environment with an occurrence rate that is not as frequent as in CC-SNe.

\begin{figure}
\begin{center}
\includegraphics[width=1.0\hsize]{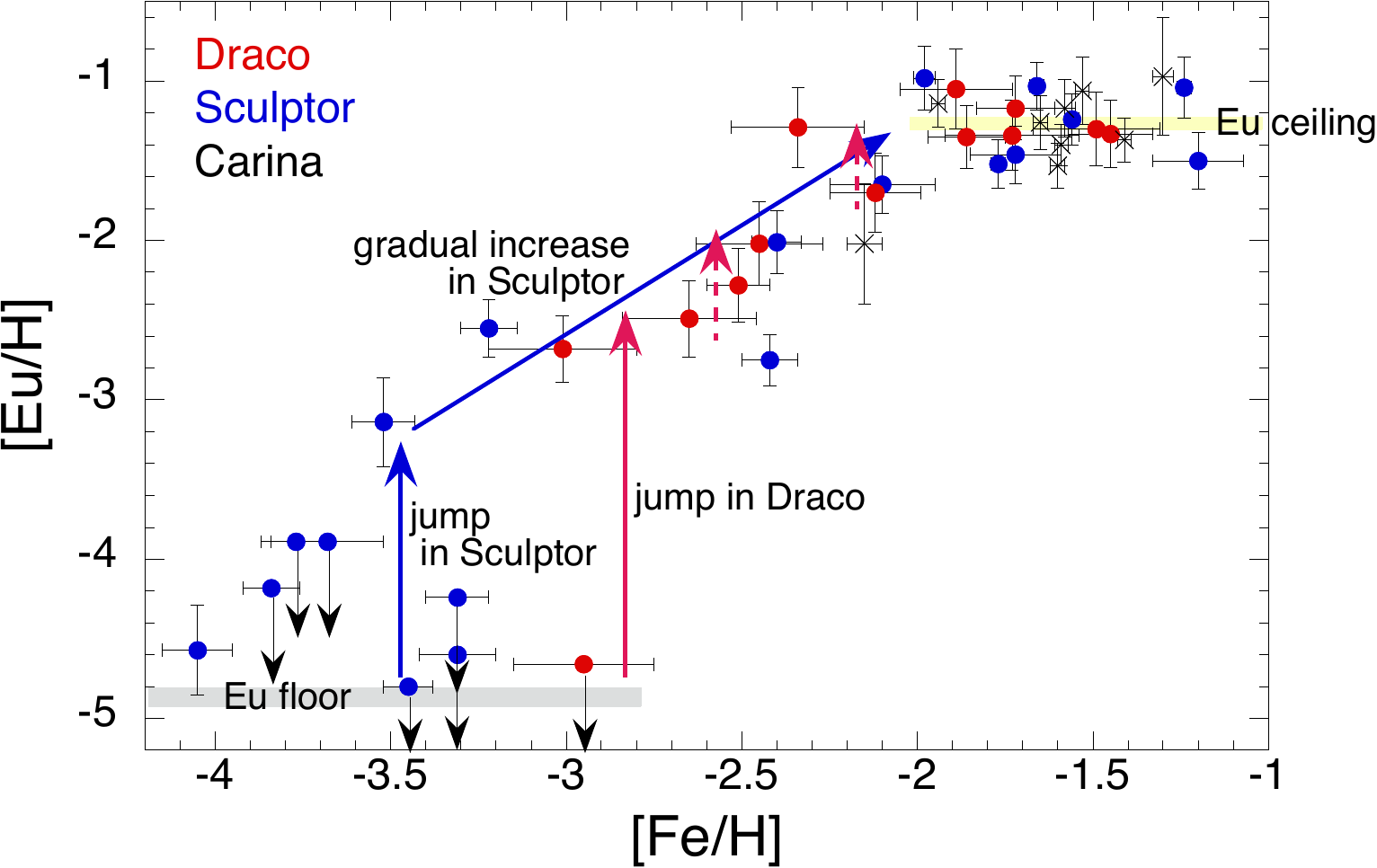}

\caption{Observed Eu feature for the Draco (red circles), Sculptor (blue circles), and Carina (crosses) dSphs. A brief illustration of early enrichment in the Draco and Sculptor dShs is shown. The observed data are updated from those of Fig.~3 in \citet{Tsujimoto_15} including [Eu/H] deduced from [Ba/H] for stars with upper limits on Eu data or no detection of Eu together with new data. The Eu floor is assigned at [Eu/H]$\sim -5$, which is implied by the Sculptor dSph. The data are originally from \cite{Shetrone_01}, \cite{Fulbright_04}, \cite{Cohen_09}, and \cite{Tsujimoto_15} for the Draco dSph; \cite{Shetrone_03}, \cite{Geisler_05}, \cite{Kirby_12}, \cite{Simon_15}, and \cite{Jablonka_15} for the Sculptor dSph; and \cite{Shetrone_03}, \cite{Lemasle_12}, and \cite{Venn_12} for the Carina dSph.
}
\end{center}
\end{figure}

Besides the mechanism of CC-SNe by neutrino heating in the standard scenario, another proposed option for r-process enrichment is an explosion triggered by fast rotations and high magnetic fields \citep[e.g.,][]{Takiwaki_09}. These magnetorotational CC-SNe (MR-SNe) can produce heavy r-process elements and thus may contribute to their enrichment during early star formation \citep{Fujimoto_08, Winteler_12, Nishimura_15, Wehmeyer_15}. In addition, it is possible to presume that the emergence of MR-SNe is inclined toward very low-metallicity stars in which the rotational velocity is expected to be high \citep[e.g.,][]{Meynet_06}.

In this Letter, we provide compelling evidence for heavy r-process enrichment by MR-SNe at an early stage of star formation inside dSphs. This Letter is organized as follows. Theoretical arguments on the chemical feature of Eu abundance in the Draco and Sculptor dSphs are first presented in Section 2. Then, the interpretation of these arguments is incorporated into the models of chemical evolution, and the results are shown in Section 3. Subsequently we discuss the astrophysical site of early Eu production in terms of nucleosynthesis in MR-SNe (Section 4) and finally conclude with Section 5. 

\section{Early Eu enrichment in the Draco and Sculptor \lowercase{d}S\lowercase{ph} Galaxies}

Chemical features of Eu abundance for faint (i.e., less massive) dSphs in the luminosity range of $3\times10^5-2\times10^6 L_\odot$, i.e., the Draco, Carina, and Sculptor dSphs, is divided into two discrete trends: (i) a remarkably increasing [Eu/H] for [Fe/H]$\ltsim-2$, and  (ii) a broadly constant [Eu/H] for [Fe/H]$\gtsim-2$. The first feature is also supported by the trend of [Ba/H] increasing versus [Fe/H] \citep{Tsujimoto_15}. It suggests that Eu production episodes occur only in the low-metallicity regime. Here we attempt to unravel how the early Eu enrichment proceeds in the Draco and Sculptor dSphs (note that the Carina dSph has only one datum for [Fe/H]$<-2$). The observed correlation of [Eu/H] with [Fe/H] is shown in Figure 1. In this figure, for the stars ([Fe/H]$<-2$) measured with only upper limits of Eu abundance or without the detection of Eu, we deduce [Eu/H] from their Ba abundances with an assumed pure r-process ratio of [Ba/Eu]=$-0.89$ \citep{Burris_00}. This procedure is validated by the fact that  the stars with [Fe/H]\ltsim $-2$ exhibit a [Ba/Eu] ratio compatible with the value assumed in the translation (i.e., [Ba/Eu]=$-0.89$) and thus are considered to be of a pure r-process origin before s-process nucleosynthesis can occur  \citep{Tsujimoto_15}.

First we see the enrichment path in the Draco. It begins with a jump of [Eu/H] from $<-4.6$ to $\sim -2.5$ between metallicities [Fe/H]$\approx -3$ and $-2.5$. To pin down the metallicity of the first Eu production episode more precisely, acquisition of more [Eu/H] data between $-3<$[Fe/H]$<-2.5$ are awaited. This includes verification of the possibility of intrinsic scatter in [Eu/H] at [Fe/H]$\approx-3$ implied by two data points. Though the subsequent path leading to a plateau of [Eu/H]$\sim -1.3$ is hard to detect, it is possible to predict the presence of a few jump-like increases in [Eu/H] as indicated by dotted arrows if we consider the errors in [Fe/H] and [Eu/H] of individual stars. In fact, such a small number of Eu production events is strongly supported by a relatively high metallicity where the first Eu jump happens. The corresponding metallicity is determined by how many CC-SNe producing Fe occur before the first event, which is directly connected to the frequency of Eu production event. For the Draco dSph with metallicities of [Fe/H]$\approx-3$ to $-2.5$, this gives about three to six events in total as shown by our calculations in the following section. Here we will deduce the actual rate of events from this total number of $3$--$6$. The stellar metallicity distribution function (MDF) for the Draco \citep{Kirby_11} indicates that $\sim 38$\% of stars populate the metallicity range of [Fe/H]$\leq-2$. Using the Kroupa IMF \citep{Kroupa_93}, we obtain the corresponding stellar mass for this metallicity range of $\sim 1.2\times 10^5$\ms \citep{Martin_08} and find that this stellar population presumably hosted $\sim600$ total CC-SNe in total. Accordingly, we conclude that r-process production events happen at a rate of one per $\sim100-200$ CC-SNe in the early evolution of Draco. In addition, we can estimate the average Eu yield of a single event. The average metallicity of [Fe/H]$\sim -1.9$ \citep{Kirby_11} is expected to result from the conversion mass fraction of the initial gas to stars of $\sim 10$\%, which is calculated by models of dwarf galaxies \citep[][see also \S 3]{Tsujimoto_11, Tsujimoto_12}. Then, the hypothesis that an initial gas cloud with a mass of $3.2\times10^6$\ms was enriched to [Eu/H]=$-1.3$ by three to six events leads to a Eu yield of $\sim(1-2)\times10^{-5}$\msp.

In the Sculptor dSph, a rise from the Eu floor happens at [Fe/H]$\sim -3.5$, a lower metallicity than Draco's, i.e., [Fe/H]\gtsim$-3$. Then, this Eu jump is followed by a continuous increase in [Eu/H] if we ignore an outlier showing a very low Eu abundance ([Eu/H]=$-2.75$) at [Fe/H]$\sim -2.4$. We note that three unpublished data sets of [Ba/H] at [Fe/H]$\approx-2.4$ \citep{Tolstoy_09} suggest $\langle {\rm [Eu/H]} \rangle=-2.2$. Such a relatively continuous feature is naturally understood if we take into consideration that the observed stellar mass of Sculptor is about eight times more than that of the Draco. We expect that the occurrence frequency of early Eu production events in the Sculptor will increase by the same factor, and as a result, more than 20 occurrences of Eu production are likely to build the smooth enrichment path as observed. Irrespective of its higher event rate, the final enrichment level should be the same as for Draco since the mass ratio of initial gas between the two dSphs is broadly equivalent to the ratio of the Eu event rate between the two. Therefore, different mass-scaled dSphs, i.e., Draco, Carina, and Sculptor, share the same plateau of [Eu/H]$\sim -1.3$. However, the evolutionary path to the plateau varies in accordance with the number of Eu production episodes.

\section{Chemical Evolution Models}

To validate the theoretical interpretation presented in the previous section, we calculate the evolution of Eu abundance for the Draco and Sculptor dSphs, adopting the following properties of Eu producers. We assume two production sites for Eu: NS mergers and one that emerges selectively in a low-metallicity regime, likely MR-SNe (see Section 4). Due to the severe rarity of NS mergers \cite[one per thousands of CC-SNe,][]{Tsujimoto_14}), the two faint dSphs did not host any NS mergers over their whole lifetimes. MR-SNe, however, contribute to Eu enrichment in the early phase with a moderately rare event rate. Here we assume that MR-SNe operate with a rate of one per 200 CC-SNe in the metallicity range of ${\rm [Fe/H]}<-2$ for the reference case. This rate gives three occurrences in Draco as already discussed, while 21 total events for Sculptor are deduced from its stellar mass and MDF \citep{Kirby_11}. In our calculations, each MR-SN is assumed to occur after every 200 CC-SNe with a progenitor metallicity lower than [Fe/H]=$-2$. For the Eu mass ejected from a single MR-SN, we adopt the yield of $2\times10^{-5}$\msp. By contrast, for the first MR-SN for both dSphs, a smaller amount of Eu mass is found to give a better fit to the observed jumps from an Eu floor, and thus $7\times10^{-6}$\ms is assumed as the Eu mass of the first MR-SN. As discussed in \S 4,  both Eu yields are within the predictions from hydrodynamical simulations of MR-SNe. In addition to the reference case, for Draco we calculate another case in which the MR-SN rate is assumed to be one per 120 CCSNe with a Eu yield of $1.2\times10^{-5}$\ms ($4\times10^{-6}$\ms for the first MR-SN).

The chemical evolution models are modified from the GCE model to account for the star formation rate (SFR) and the IMF as done by \citet{Tsujimoto_12}. To reproduce the observed MDFs with a mean [Fe/H] of $-1.93$ and $-1.68$ for the Draco and Sculptor dSphs, respectively \citep{Kirby_11}, we adjust the SFR together with a steep IMF ($x$=$-1.6$) as in the Large Magellanic Cloud. Here we assume that the period of star formation continues for 4 Gyr for Draco \citep[cf.][]{Dolphin_02} and 6 Gyr for Sculptor \citep{Boer_12}, respectively. The initial Eu abundance (i.e., the Eu floor) is set to be [Eu/H]=$-5$. The results from our models are shown in Figure 2.

\begin{figure}
\begin{center}
\includegraphics[width=1.0\hsize]{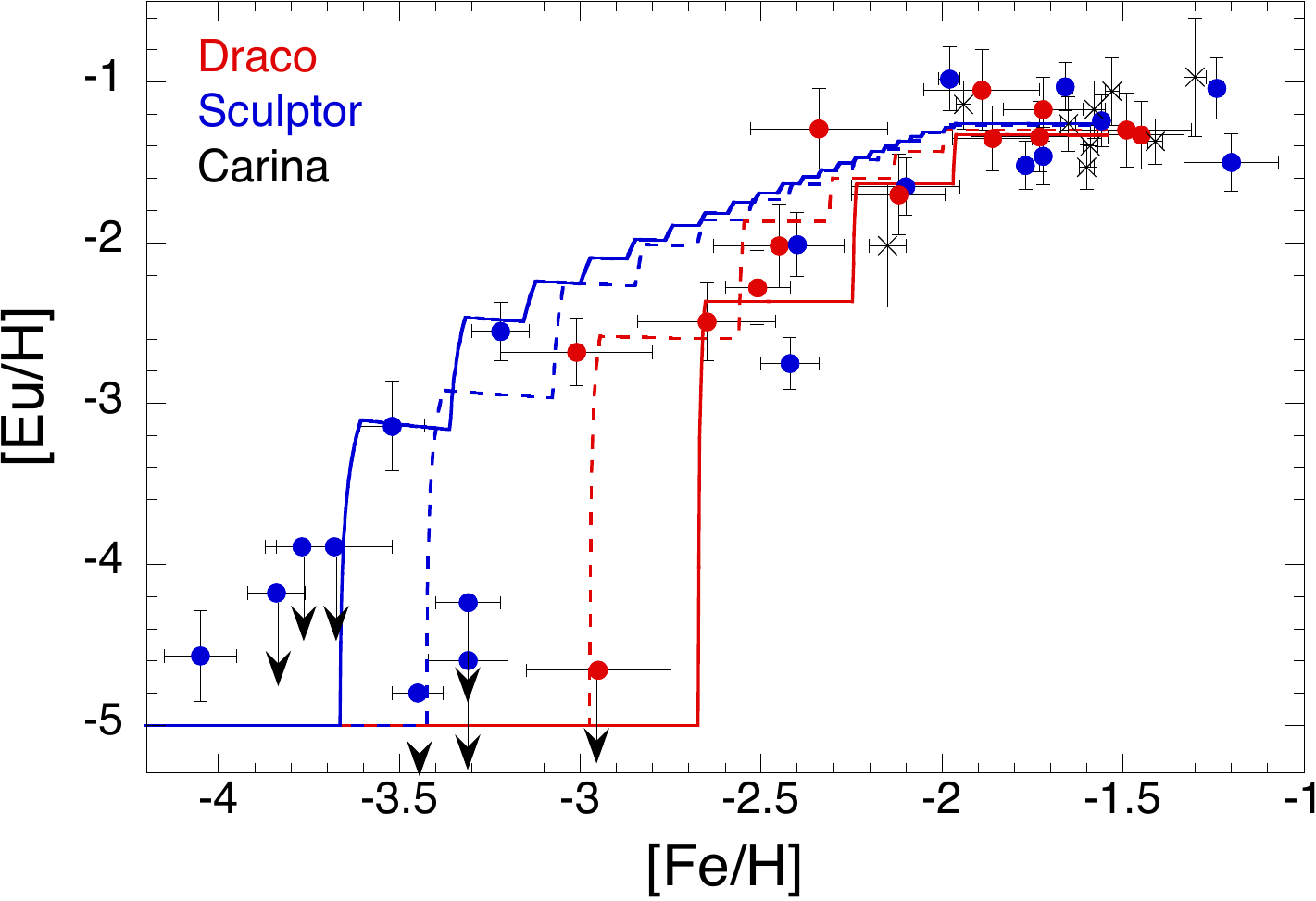}
\caption{Predicted evolutions of Eu abundance for the Draco (red curves) and the Sculptor (blue curve) dSphs, compared with the observed data shown in Fig.~1. The Eu production events are assumed to occur at a rate of one per 200 CC-SNe (solid curves) or one per 120 (350) CC-SNe (dashed red (blue) curve) for ${\rm [Fe/H]}<-2$ while staying dormant for ${\rm [Fe/H]}>-2$. A resultant smaller event rate (three: solid curve; five: dashed curve) in the Draco is attributed to its lower stellar mass compared with the Sculptor. The pre-enriched level is set at ${\rm [Eu/H]} =-5$.
}
\end{center}
\end{figure}

The observed values of [Fe/H] corresponding to the jump from a pre-enriched level of [Eu/H] for the two dSphs are well reproduced by our models. The higher jump of [Eu/H] at higher [Fe/H] in Draco can be interpreted as a result of a lower frequency of MR-SNe owing to a smaller total stellar mass than Sculptor. The subsequent ladder-like increasing feature of [Eu/H] in Draco is also in good agreement with observations. On the other hand, a gradually increasing trend after the first jump until the plateau-like abundance (i.e., [Eu/H]$\sim -1.3$) is predicted by the model for Sculptor. Three data points giving the low upper limits of [Eu/H] between $-3.5<$[Fe/H]$<-3.3$ suggest a more delayed first jump. Accordingly, the result of another model in which 12 total events are assumed to occur with a rate of one per 350 CCSNe is shown by the dashed blue curve. The agreement of our models with observations for both Draco and Sculptor suggest that Eu production episodes occur more frequently than NS mergers, but much less frequently than CC-SNe and with a high Eu yield.

\section{Nucleosynthesis in MR-SNe}

\begin{figure}
\begin{center}
\includegraphics[width=1.0\hsize]{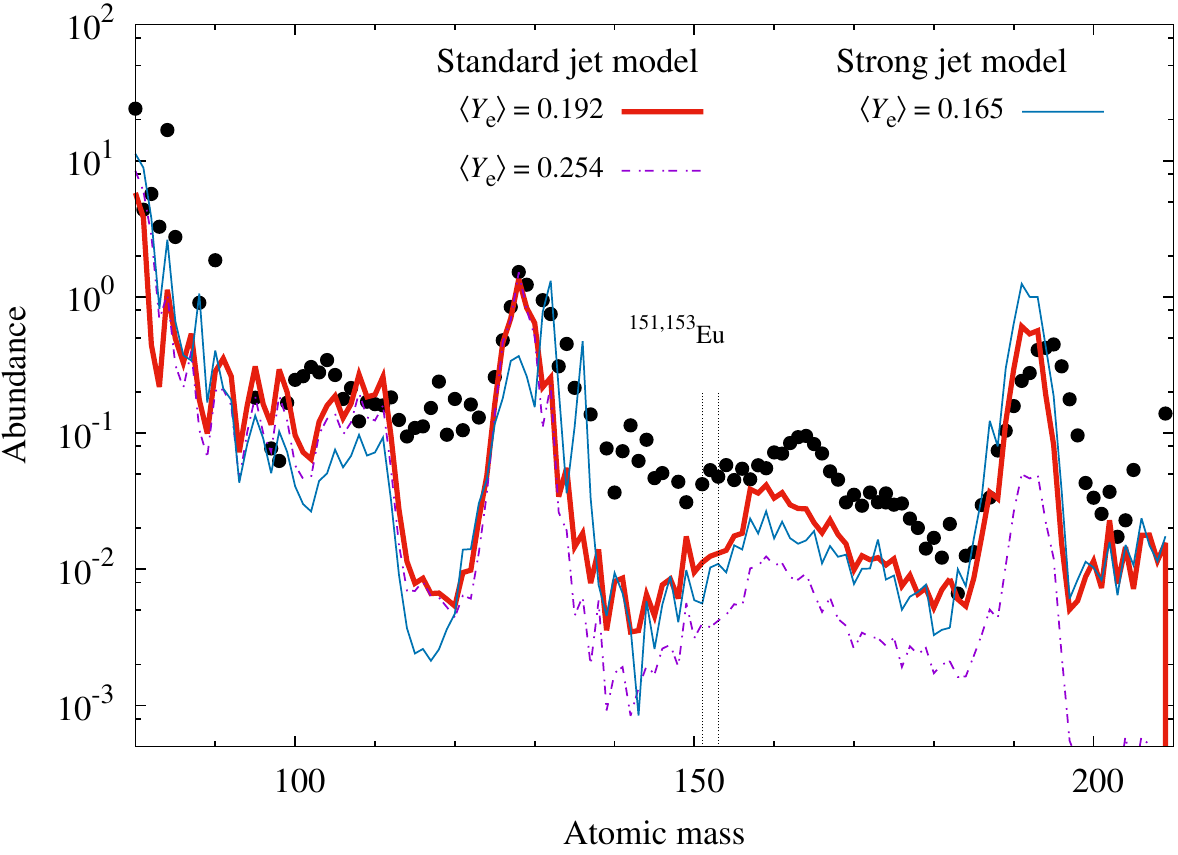}
\caption{Nucleosynthesis abundance patterns predicted by MR-SN models for three different cases, compared with the solar abundance pattern \citep{Arlandini_99}. See the text for the nucleosynthesis models.
}
\end{center}
\end{figure}

As discussed in the previous sections, the early chemical evolution of the Draco and Sculptor dSphs demands a large amount of Eu production such as $\sim (1-2)\times 10^{-5} M_\odot$ per single event. This value far exceeds the assumed amount ($10^{-7}-10^{-6}$\msp) from the ejecta of canonical CC-SNe with neutrino-driven winds in GCE models \citep[e.g.,][]{Argast_04}, while it is smaller than the expectation from an NS merger ($\sim 10^{-4}$\msp). Here we claim that the implied Eu yields agree well with the prediction from MR-SNe, based on the recent calculations by \citet{Nishimura_15}. In their models, a typical case (i.e., {\tt B11$\beta$1.00}) results in a Eu mass $M({\rm Eu}) = 0.724\times10^{-5}$\ms in the jet-like ejecta and a total mass of r-process nuclei of $1.53\times 10^{-2} M_\odot$. Inside the ejecta, the values of $Y_{\rm e}$ vary between 0.15 and 0.5 and an average in the region where r-process operates (i.e., $Y_{\rm e}<0.3$) is $\langle Y_{\rm e} \rangle=0.192$. The resultant nucleosynthesis pattern is shown by the red curve in Figure 3, compared with the observed solar abundance pattern \citep[filled circles:][]{Arlandini_99}. In this calculation, we used reaction rates based on the finite-range droplet mass model which is available in the reaction rate library REACLIB \citep{Rauscher_00}, leading to a relatively low Ba/Eu ratio of [Ba/Eu]$=-1.44$. However, a recent improvement \citep{Moller_12} indicates that the production of isotopes in the region including the rare earth peak has increased by a few factors \citep{Kratz_14}, which may lead to a match with the solar pattern, giving $M({\rm Eu}) = 1.63 \times 10^{-5} M_\odot$ and a [Ba/Eu] ratio close to [Ba/Eu]=$-0.89$. In addition, \citet{Nishimura_15} pointed out that the total mass of ejecta can be larger by a few factors in more energetic jet-like explosion models. This strong jet model gives ${\rm [Ba/Eu]}=-0.55$ (the blue curve).  Accordingly, the Eu mass of $\sim (1-2)\times 10^{-5} M_\odot$ is indeed within the prediction by MR-SN models.

Besides the above jet-like explosions, another scenario that would eject a significant amount of Eu is from MR-SNe,
which release a massive amount of moderate neutron-rich ejecta. According to recent updated magnetohydrodynamic simulations that resolve the magnetorotational instability \citep{Sawai_14}, the ejecta due to neutrino heating enhanced by fast rotation and large magnetic fields have much lower values of $Y_{\rm e}$ for the cases of explosion without jet-like outflows. As a result, the ejecta are a moderate neutron-rich state. To obtain the  nucleosynthesis results for the corresponding case, we set higher $Y_{\rm e}$ values inside the ejecta in the jet model (see Section 5.2 of Nishimura et al. 2012) and obtain results for $\langle Y_{\rm e} \rangle = 0.254$, as shown by the dashed curve in Figure 3. The ejecta mass including the total r-process nuclei of $2.15 \times 10^{-2} M_\odot$ \citep{Nishimura_15} gives $M({\rm Eu})=0.85\times 10^{-5} M_\odot$, which is broadly equivalent to the mass predicted by the jet-like explosion models.

\section{Conclusions}

We show that the properties of r-process production events that are important for early star formation,  i.e., the  frequency and yield,  can be assessed from the recently revealed Eu feature of two less massive dSphs. The frequency is estimated to be about one per $100$--$200$ CC-SNe in a low-metallicity regime, while this frequency decreases to a level similar to or less than that of NS mergers for ${\rm [Fe/H]}>-2$. We identify high frequency events at low-metallicity together with the implied high Eu yield as MR-SNe. Accordingly, we have the following picture of r-process enrichment among galaxies. In dwarf galaxies such as the Draco and Sculptor dSphs, MR-SNe at an early phase are the only contributors of the nucleosynthesis of heavy r-process elements. However, in more massive galaxies, including our own, in addition to MR-SNe with their metallicity-dependent frequencies, NS mergers enrich the ISM with heavy r-process elements over the whole lifetime of a galaxy at a rate of one per thousands of CC-SNe. Since enrichment by MR-SNe is expected to start much earlier than NS mergers, the presence of extremely metal-poor stars enriched by r-process elements as observed in the Galactic halo is naturally explained by our scheme.

We predict that the frequency of MR-SNe is fairly small in the relatively metal-rich galaxies. Thus their occurrences should be detected with a high rarity in the present-day Universe. MR-SNe are likely to be  associated with magnetars, the rate of which is estimated to be $0.003-0.06\ {\rm yr}^{-1}$ in the Galaxy \citep{Muno_08}. We accordingly anticipate that a few percent or less of magnetars are the end result of MR-SNe. A link between MR-SNe and rare phenomena such as superluminous SNe  \citep{Dessart_12} and ultra-long GRBs \citep{Greiner_15}, which may be energized by magnetars, is worth investigating in future work.

\acknowledgements

This work was supported by JSPS KAKENHI Grant Number 15K05033. N.N. was financially supported by the ERC (EU- FP7-ERC-2012-St Grant 30690). The authors thank A. Cristini for proofreading the manuscript. Parts of the numerical computations were carried out on computer facilities at CfCA in NAOJ and the COSMOS (STFC, DiRAC Facility) at DAMTP at the University of Cambridge.

\end{document}